\documentclass[11pt,a4paper,onecolumn,reqno]{amsart}
\usepackage[a4paper, margin=2.3cm, bmargin=3cm]{geometry}

\usepackage{graphicx,stfloats} \usepackage{color}
\usepackage{amsmath}
\usepackage{amsaddr}
\usepackage{bm, soul}
\usepackage{mathtools}
\usepackage[colorinlistoftodos,prependcaption,textsize=tiny]{todonotes}
\usepackage{braket}
\usepackage{hyperref}
\usepackage[square,comma,sort&compress, numbers]{natbib}
\usepackage{enumerate}

\usepackage{tikz}
\usepackage{pgfplots}
\pgfplotsset{compat=1.9}
\usetikzlibrary{calc,arrows,fadings,decorations.pathreplacing,decorations.markings,patterns,shapes.geometric}
\tikzset{>=stealth,inner sep=0pt, outer sep=2pt,}
\tikzset{vecteur/.style={->,thick,color=black,smooth}}

\usepackage{placeins}

\renewcommand{\st}[1]{}

\usepackage[version=3]{mhchem} \usepackage{siunitx}
 \usepackage{multirow}

\usepackage{etoolbox}
\patchcmd{\pprintMaketitle}
  {\hrule\vskip12pt}
 {\hrule\vskip12pt\ifvoid\extrainfobox\else\unvbox\extrainfobox\par\vskip12pt\fi}
 {}{}

\newsavebox\extrainfobox

\usepackage{caption}
\title{Flame-vortex interaction during turbulent side-wall quenching and its implications for flamelet manifolds}

\author[stfs]{Matthias Steinhausen$^{1,*}$, Thorsten Zirwes$^{2, 3}$, Federica Ferraro$^{a}$, Arne Scholtissek$^{1}$, Henning Bockhorn$^{2}$, Christian Hasse$^{1}$}
\email{steinhausen@stfs.tu-darmstadt.de} 
\address[]{$^1$Technical University of Darmstadt, Department of Mechanical Engineering, Simulation of reactive Thermo-Fluid Systems, Otto-Berndt-Stra{\ss}e 2, 64287 Darmstadt, Germany\\
$^2$Engler-Bunte-Institute, Karlsruhe Institute of Technology, Engler-Bunte-Ring 7, 76131 Karlsruhe, Germany\\
$^3$Steinbuch Centre for Computing, Karlsruhe Institute of Technology, Hermann-von-Helmholtz-Platz 1, 76344 Eggenstein-Leopoldshafen, Germany
}

\begin{document}
\pagestyle{plain}

\maketitle

\begin{abstract} 
  In this study, the thermochemical state during turbulent flame-wall interaction of a stoichiometric methane-air flame is investigated using a fully resolved simulation with detailed chemistry. The turbulent side-wall quenching flame shows both head-on quenching and side-wall quenching-like behavior that significantly affects the CO formation in the near-wall region. The detailed insights from the simulation are used to evaluate a recently proposed flame (tip) vortex interaction mechanism identified from experiments on turbulent side-wall quenching. It describes the entrainment of burnt gases into the fresh gas mixture near the flame's quenching point. The flame behavior and thermochemical states observed in the simulation are similar to the phenomena observed in the experiments. A novel chemistry manifold is presented that accounts for both the effects of flame dilution due to exhaust gas recirculation in the flame vortex interaction area and enthalpy losses to the wall. The manifold is validated in an \textit{a-priori} analysis using the simulation results as a reference. The incorporation of exhaust gas recirculation effects in the manifold leads to a significantly increased prediction accuracy in the near-wall regions of flame-vortex interactions.
\end{abstract}


\section{Introduction} \addvspace{10pt}
Technical combustors for power generation, such as internal combustion engines or gas turbines, are typically operated under turbulent flow conditions and enclosed by walls. The turbulent combustion process in the combustion chamber is a complex multi-scale, multi-physics phenomenon that still poses a challenge for numerical simulations. During turbulent flame-wall interactions (FWIs), the complexity increases even further. In technical systems, the temperature of the combustor walls is often lower than the gas temperature. In these cases, the flame is affected by enthalpy losses at the walls, leading to incomplete combustion, which lowers the overall efficiency and enhances pollutant formation~\cite{Poinsot2005a}. Additionally, FWI can lead to undesired flame behavior, such as flame flashback~\cite{Fritz2004}.

Two major effects need to be considered to simulate turbulent FWI processes: (i) the fluctuations of the reactive scalars and (ii) the influence of the cold walls on the combustion chemistry. In Direct Numerical Simulations (DNS) of FWIs~\cite{Gruber2010, Ahmed2021, Steinhausen2022, Jiang2021a}, all relevant scales of transport and finite-rate chemistry are resolved, resulting in high computational costs that render this approach unfeasible for the simulation of real combustion applications. In the simulation of practical systems, Reynolds-Averaged Navier-Stokes (RANS) and Large Eddy Simulations (LESs) are typically used, which require a suitable turbulence chemistry interaction (TCI) closure model and a reduction of the combustion chemistry. While TCI closure approaches for FWI have been addressed recently~\cite{Steinhausen2022}, this study focuses on the latter using chemistry manifolds~\cite{VanOijen2000, Gicquel2000, Maas1992, Bykov2007} that combine the high prediction accuracy of the thermochemical state of a finite-rate chemistry simulation with low computational costs. 

Chemistry manifolds for FWI have been developed and validated in multiple studies of laminar flames~\cite{Ganter2018, Strassacker2021, Steinhausen2021, Efimov2019b} against fully resolved finite-rate chemistry simulations and measurements of the near-wall thermochemical states~\cite{Kosaka2018, Zentgraf2021a}. 
These studies showed that the near-wall flame structure cannot be fully captured by 1D laminar flamelets based on burner-stabilized flames. However, transient head-on quenching (HOQ) flames are suitable to describe the flame structure during FWI. It was validated in quenching laminar flames~\cite{Ganter2018, Strassacker2021, Steinhausen2021, Efimov2019b} and is extended to turbulent flames in the present work.
Recent numerical~\cite{Palulli2019, Jiang2021a} and experimental~\cite{Zentgraf2021} studies also show a high dependency of the thermochemical state in transient and turbulent FWIs caused by velocity perturbations interacting with the quenching flame. Palulli et al.~\cite{Palulli2019} investigated the influence of velocity perturbations on the near-wall thermochemical state, more specific the local heat-release rate and the CO formation. In that work, a 2D finite-rate chemistry simulation of a side-wall quenching (SWQ) flame was performed that is prone to velocity perturbations of varying frequencies. While at low and high forcing frequency, only SWQ-like behavior was observed, at an intermediate frequency, the flame showed both head-on quenching (HOQ) and SWQ-like behavior that significantly affects the CO formation at the wall. In~\cite{Jiang2021a} a 3D DNS of a diluted methane-air flame undergoing SWQ was performed and the CO formation in the flame was analyzed. In the study, the thermochemical state of the DNS was compared to 1D freely propagating flames with different amounts of exhaust gas added to the fresh gas mixture and opposed-flow flames with different strain rates. In the core flow, a good agreement of the thermochemical states was observed. In close vicinity to the wall, however, the thermochemical state was not captured by the adiabatic flame configurations. Finally, Zentgraf et al.~\cite{Zentgraf2021} investigated the thermochemical state during turbulent atmospheric SWQ using simultaneous quantitative measurements of $\mathrm{CO}$, $\mathrm{CO_2}$ and temperature. They demonstrated that the observed thermochemical states in the turbulent SWQ scenario differ significantly from the laminar configuration investigated in~\cite{Zentgraf2021a}. In accordance to Palulli et al.~\cite{Palulli2019} HOQ and SWQ-like behavior was observed in the turbulent flames. Furthermore, the authors proposed a possible flame-vortex interaction (FVI) mechanism that explains the differences between the observed laminar and turbulent states as originating from the recirculation of burnt exhaust gases into the unburnt part of the flame.

In the present study a three-dimensional fully-resolved simulation with detailed chemistry of the FWI of a stoichiometric methane-air flame in a fully developed turbulent channel flow is performed. The objective of this work is threefold: 
\begin{enumerate}[(i)]
    \item to investigate the FVI mechanism proposed in~\cite{Zentgraf2021} for a turbulent SWQ with a focus on the observed thermochemical states. In this context, the FVI mechanism is numerically verified and analyzed based on a time series of the performed simulation;
    \item to model the effects of FVI based on the gained numerical insights using a novel chemistry manifold that accounts for the effects of exhaust gas recirculation (EGR) and enthalpy losses at the wall;
    \item to validate the manifold in an \textit{a-priori} manner using the detailed simulation results as a reference.
\end{enumerate}

\section{Numerical setup} \addvspace{10pt}
\subsection{Turbulent side-wall quenching flame}
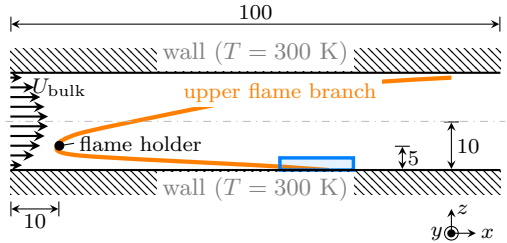
\begin{figure}[!htb]
    \centering
    \begin{tikzpicture} [scale=6.77/10.5]
     	\definecolor{azure}{rgb}{0.0, 0.5, 1.0}
    	\foreach \y in {-0.9,-0.7,-0.5,-0.3,-0.1,0.9,0.7,0.5,0.3,0.1}
    	\draw[vecteur] (0,\y)--++({2*(1-(\y*\y)^(1/2))^(1/7)-1.2},0);
        \draw (1,0.7)node{\scriptsize $U_\text{bulk}$};
    
        \draw[|<->|] (0,2.0)--++(10,0) node[midway, above]{\scriptsize 100};
        \draw[|<->|] (9,-1)--++(0,1) node[midway, right]{\scriptsize 10};
        
        \draw[ultra thick, orange] plot [smooth] coordinates { (9,0.9) (7, 0.78) (6, 0.65) (4.5, 0.4) (3, 0.1) (1.4, -0.25) (0.95, -0.5) (1.4, -0.7) (5, -0.9) (5.8, -0.95) (6.5, -1) };
        \node[orange, inner sep=2pt, fill=white, anchor=center] at (5.5, 0.6) {\scriptsize upper flame branch};
    
    	\filldraw (1,-0.5) circle (0.09);
        \draw[-] (1, -0.5) -- (1.3,-0.45) node[right]{\scriptsize flame holder};
        \draw[|<->|] (0,-1.8)--++(1,0) node[midway, below]{\scriptsize 10};
        \draw[|<->|] (8,-1)--++(0,0.5) node[midway, right]{\scriptsize 5};

        \draw[->] (9.0,-2.3)--++(0,0.5) node[above, right]{\scriptsize $z$};
        \draw[->] (9.0,-2.3)--++(0.5,0) node[above, right]{\scriptsize $x$};
        \node[anchor=center,draw,circle] at (9.0, -2.3) {\tiny \textbullet};
        \node[anchor=center] at (8.7, -2.3) {\scriptsize $y$};
        
        \draw[line width=0.45mm, azure, fill=azure, opacity=0.1]  (5.5,-1) rectangle (7,-0.75);
        \draw[line width=0.45mm, azure]  (5.5,-1) rectangle (7,-0.75);

        \draw[thick](0,-1)--++(10,0);
    	\fill[pattern=north west lines] (0,-1.5) rectangle (10,-1);
    	\draw[thick](0,1)--++(10,0);
    	\fill[pattern=north west lines] (0,1) rectangle (10,1.5);
    	
        \draw[lightgray, dash dot] (-0.1,0)--++(10.2,0) node[right]{};

        \node[gray, inner sep=2pt, fill=white, anchor=center] at (5, -1.4) {\footnotesize wall ($T=300$ K)};
        \node[gray, inner sep=2pt, fill=white, anchor=center] at (5, 1.4) {\footnotesize wall ($T=300$ K)};
    
    \end{tikzpicture}
    \vspace{1 pt}
   \caption{Schematic of the numerical setup. In the (statistically independent) lateral direction ($y$) the channel width is 3~cm. The region of interest analyzed in Figs.~\ref{fig:time-series} and~\ref{fig:validate-lookup} is shown as a blue rectangle. All measurements are given in mm.}
  \label{fig:DNS-setup}
\end{figure}

In the following, the numerical setup of the turbulent SWQ flame analyzed in this work is outlined. The setup is inspired by Gruber et al.~\cite{Gruber2010} and was used for the analysis of TCI closure in~\cite{Steinhausen2022}. Figure~\ref{fig:DNS-setup} shows a schematic view of the numerical setup. In a fully developed turbulent channel flow, a V-shaped premixed stoichiometric methane-air flame is stabilized at a flame holder and undergoes side-wall quenching at the (cold) channel walls. The wall temperature is fixed and equal to the inflow temperature $T_\text{in}=T_\text{wall}=300~\text{K}$. The channel flow has a Reynolds number of $Re=(U_\text{bulk}H)/\nu\approx2770$, with $U_\text{bulk}$ being the mean flow velocity, $H$ the channel half-width and $\nu$ the kinematic viscosity. The mean inflow velocity is $U_\text{bulk}=4.4~\text{ms$^{-1}$}$. The flame holder is modeled as a cylindrical region of burnt gas temperature with the center at $H/2$ above the bottom wall. 
Note that the equivalence ratio and wall temperatures are chosen to allow a comparison of the physical phenomena with recent experimental investigations~\cite{Jainski2017a, Zentgraf2021}. While the work of Zentgraf et al.~\cite{Zentgraf2021} focuses on dimethyl ether flames, a methane-air flame was also investigated in the experimental campaign. In their study, the equivalence ratio of the dimethyl ether flame was chosen to match the flame speed of a stoichiometric methane-air flame, which is investigated in this study.

A \textit{non-reactive simulation} of the turbulent channel flow is performed to generate appropriate inflow conditions for the reactive case. The channel dimensions of the non-reactive case are $x\times y \times z=14H \times 3H \times 2H$, with $x$ being the stream-wise, $y$ the lateral, and $z$ the wall-normal direction. In the stream-wise and lateral direction, periodic boundary conditions are applied, while the wall is modeled with a no-slip boundary condition. The computational grid is refined towards the walls with a wall resolution of $25~\text{\textmu m}$ ($y^+=0.24$), resulting in a total of 60 million hexahedral cells. The inflow velocity fields at the boundary serve as inflow conditions for the reactive case and are stored with a time-step of $\Delta t=3~\text{\textmu s}$.

The \textit{reactive simulation} has a reduced channel dimension in stream-wise direction of $10H$. The purely hexahedral, orthogonal mesh is refined towards the bottom wall with a wall resolution of $12~\text{\textmu m}$ ($y^+=0.14$) and a total of 200 million cells is used, ensuring a sufficient grid resolution (Kolmogorov length scale $\eta>45~\text{\textmu m}$, laminar flame thickness $\delta_L = \left( T_\text{burnt} - T_\text{unburnt} \right) / \left| \delta T / \delta x \right|_\text{max} \approx 0.5~\text{mm}$). Note that the resolution at the bottom wall is motivated by the flame-wall interaction zone, where the flame can move as close as $100~\text{\textmu m}$ towards the cold wall~\cite{Zirwes2021}. The boundary conditions of the domain are set as follows: At the inlet, the inflow velocity fields from the non-reactive simulation are employed. Therefore, the velocity fields are spatially and temporally interpolated to the inlet boundary face at every time step of the reacting simulation. In the lateral direction, periodic boundary conditions are applied, while the walls are modeled using a no-slip boundary condition for the velocity, a zero-gradient boundary condition for the species, and a fixed temperature boundary condition ($T_\text{wall}=300~\text{K}$). Finally, at the outlet, a Dirichlet boundary condition is employed for the pressure, while a zero-gradient boundary condition is used for the reactive scalars and the velocity. The source terms are described using a reduced version of the CRECK mechanism~\cite{Ranzi2012} that consists of 24 species and 165 reactions and a unity Lewis number assumption for the molecular diffusion coefficients. Table~\ref{tab:setup} summarises the most important setup parameters.

\begin{table}[h!] \small
  \centering
  \caption{Numerical setup of the reactive case}
  \centerline{\begin{tabular}{ll}
      \hline
      \textbf{Parameter} & \textbf{Property} \\
      \hline
      Gas mixture & Methane-air ($\Phi=1$) \\
      Reaction mechanism & Reduced CRECK~\cite{Ranzi2012} \\
      Diffusion model & Le=1 transport \\
      ($x \times y \times z$) & ($100 \times 30 \times 20$)~mm \\
      Anchor position ($x$, $z$) & ($10~\text{mm}$, $5~\text{mm}$) \\
      Anchor radius & $0.9~\mathrm{mm}$ \\
      Bulk velocity & $4.4~\mathrm{ms^{-1}}$ \\
      Inlet temperature & $300~\text{K}$ \\
      Wall temperature & $300~\text{K}$ \\
      Reynolds number & $2770$ \\
      \hline
  \end{tabular}}
  \label{tab:setup}
\end{table}

The simulations are performed with an in-house solver~\cite{Zirwes2018, Zirwes2018improved} that uses finite-rate chemistry. In~\cite{Zirwes2018} the solver was validated to be suitable for DNS-like simulations using multiple DNS reference cases from the literature. The spatial discretization is of fourth-order, and a second-order fully implicit backward scheme is used for the temporal discretization. The reactive simulation was performed on 32678 CPU cores (AMD EPYC 7742), and more than 18 million core-h have been consumed.

\newpage

\subsection{Laminar side-wall quenching flame}
In addition to the turbulent case, a corresponding laminar SWQ simulation is performed. The setup of the laminar case is similar to the one presented in~\cite{Ganter2018, Steinhausen2022}. The simulation is performed in a two-dimensional domain, where a flame is stabilized at the inlet away from the wall by injecting hot exhaust gases at equilibrium conditions. The flame approaches the wall with a wall temperature of 300~K, where it undergoes SWQ. The numerical solver employed is similar to the turbulent simulation described above. The boundary conditions and the numerical grid of the simulations are equal to the one described in~\cite{Steinhausen2022} except for the wall temperature.


\section{Thermochemical state in turbulent SWQ} \addvspace{10pt}
In this section, the thermochemical state of the turbulent SWQ simulation is analyzed in the near-wall region up to a normalized wall distance of $z/\delta_\text{L}=6$ and compared to the thermochemical states of laminar SWQ and a freely propagating (fp) flame for reference. 

\begin{figure*}[!htb]
    \centering
    \includegraphics[scale=1.0]{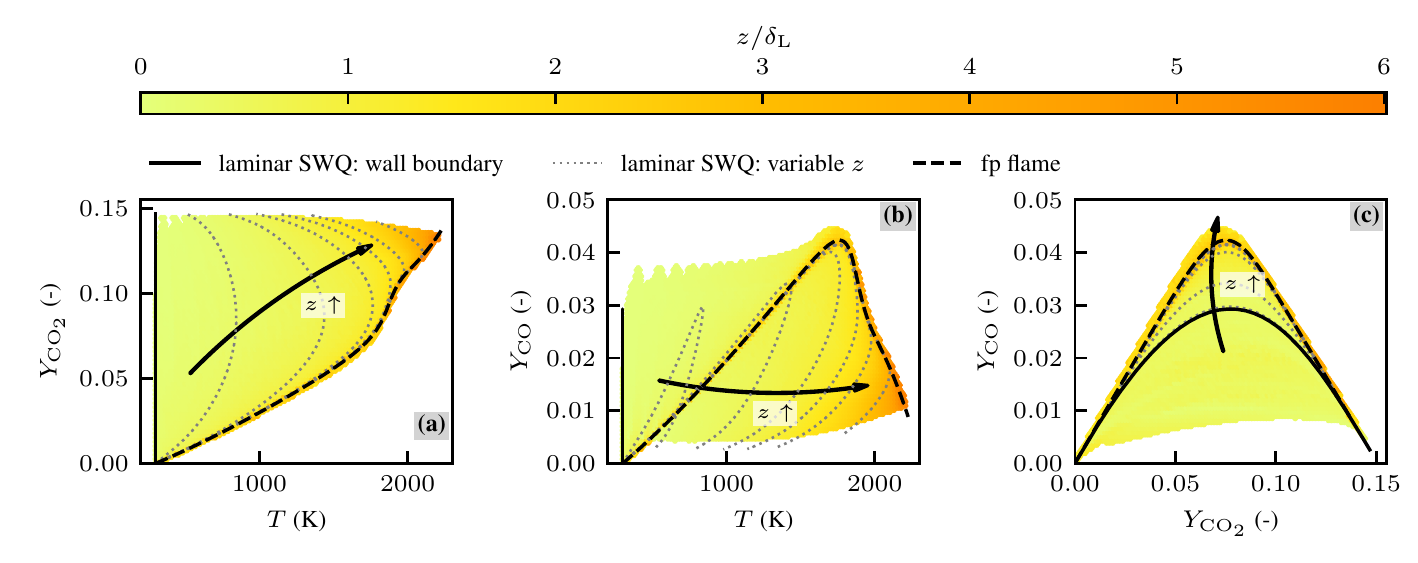}
    \vspace{5pt}
    \caption{Thermochemical states in the turbulent SWQ simulation colored by wall distance $z$. For reference, a 1D freely propagating (fp) flame and lines extracted parallel to the wall from a laminar SWQ are shown.}
    \label{fig:thermochemical-state}
\end{figure*}

Figures~\ref{fig:thermochemical-state}(a-c) show the thermochemical states from the turbulent SWQ represented by temperature $T$ and $\mathrm{CO_2}$ and $\mathrm{CO}$ mass fractions. The data is colored by the normalized wall distance $z/\delta_\text{L}$. In addition to the turbulent SWQ, a corresponding laminar SWQ is shown. The spanned thermochemical state is bounded by the cold wall boundary (solid black line) and a freely propagating flame (dashed black line). Furthermore, different lines extracted parallel to the wall are shown to visualize the influence of the wall distance on the thermochemical state (gray dotted lines).

In the $T$-$Y_\mathrm{CO_2}$ plane (Fig.~\ref{fig:thermochemical-state}(a)), the state space spanned by the turbulent SWQ is also fully covered by its laminar counterpart. Similar observations can be made for $T$-$Y_\mathrm{CO}$ in Fig.~\ref{fig:thermochemical-state}(b). Only for low temperatures, the turbulent case shows states with higher $Y_\mathrm{CO}$ compared with its laminar counterpart. Finally, the $Y_\mathrm{CO_2}$-$Y_\mathrm{CO}$ plane (Fig.~\ref{fig:thermochemical-state}(c)) is addressed. In the laminar SWQ, $Y_\mathrm{CO}$ shows a conditional maximum and minimum for a given value of $Y_\mathrm{CO_2}$
\begin{align}
    \max \left( Y_\mathrm{CO} | Y_\mathrm{CO_2} \right) &= \left( Y_\mathrm{CO} | Y_\mathrm{CO_2} \right)_\text{fp flame} \ , \\
    \min \left( Y_\mathrm{CO} | Y_\mathrm{CO_2} \right) &= \left( Y_\mathrm{CO} | Y_\mathrm{CO_2} \right)_\text{$z$ = 0~mm} \ ,
\end{align}

\noindent with $\left(Y_\mathrm{CO}|Y_\mathrm{CO_2}\right)$ being the value of $Y_\mathrm{CO}$ for a given value of $Y_\mathrm{CO_2}$. While the highest $\left(Y_\mathrm{CO}|Y_\mathrm{CO_2}\right)$ is present in the undisturbed part of the flame (freely propagating flame / high wall distance), $\left(Y_\mathrm{CO}|Y_\mathrm{CO_2}\right)$ decreases with the wall distance and reaches a minimum at the wall ($z=0~\text{mm}$), where the flame is quenched and cooled down to wall temperature. In the turbulent SWQ, the minimum value of  $\left( Y_\mathrm{CO} | Y_\mathrm{CO_2} \right)$ is lower compared to the laminar counterpart, showing values comparable to the unburnt and burnt gas state over the whole range of $Y_\mathrm{CO_2}$. A flame-vortex interaction (FVI) mechanism was proposed in~\cite{Zentgraf2021} that explains the differences in the observed states of the laminar and turbulent SWQ originating from the mixing of (cold) burnt products with unburnt gases in the close vicinity of the quenching point.

\begin{figure*}[!htb]
    \vspace{-5pt}    
    \centering
    \includegraphics[scale=1.0]{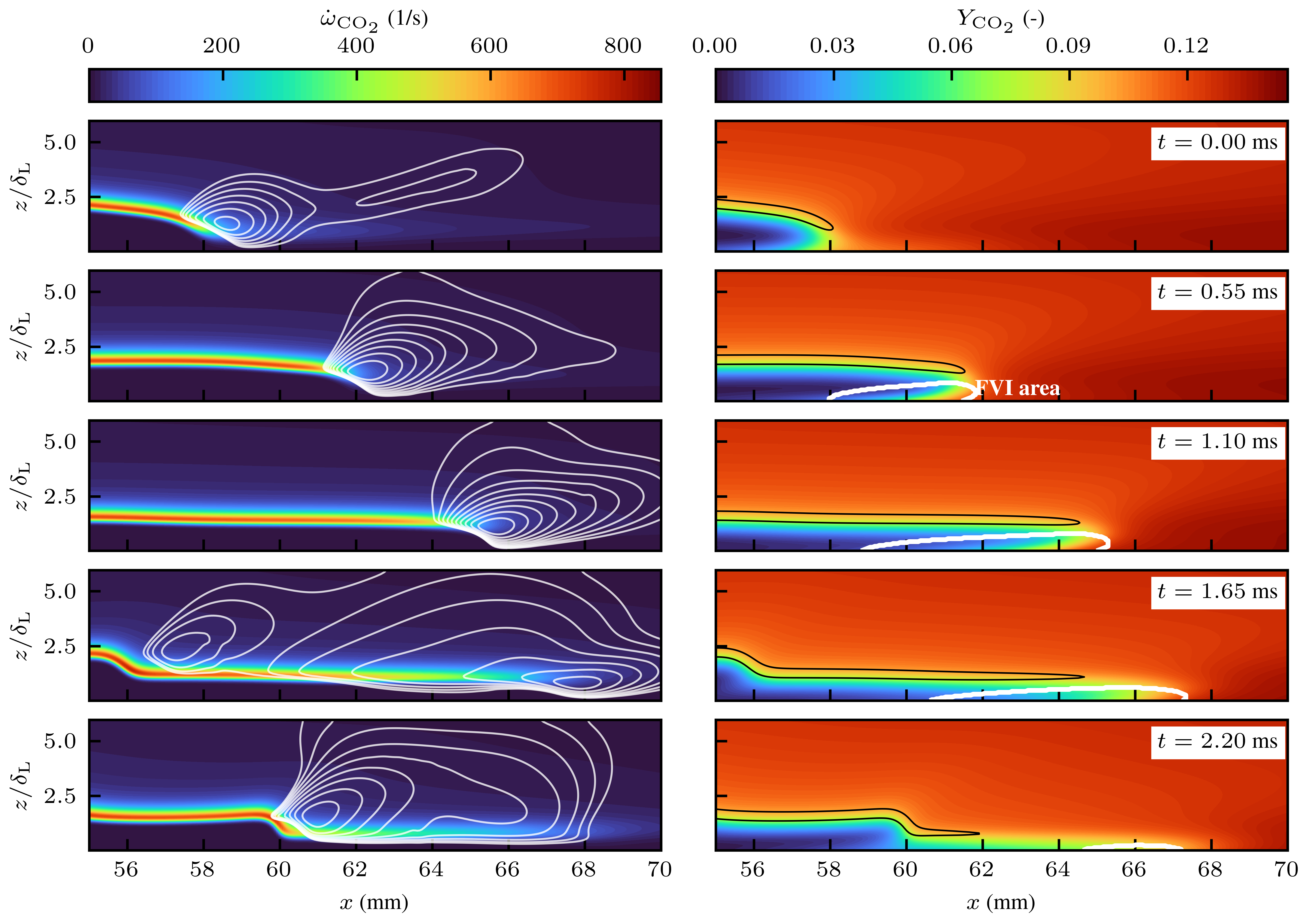}
    \vspace{5pt}
    \caption{Time series of a slice in lateral direction through the turbulent flame. On the left, the reaction rate of $\mathrm{CO_2}$ is shown together with vortical structures visualized by the Q-criterion (white lines). On the right, $Y_\mathrm{CO_2}$ is shown and the flame front is visualized by a contour of $\dot{\omega}_\mathrm{CO_2}$ (black). Additionally, the white isocontour represents the area of FVI. The area shown in the slices is also depicted Fig.~\ref{fig:DNS-setup} as a blue rectangle.}
    \label{fig:time-series}
    \vspace{-5pt}    
\end{figure*}

\begin{figure}[!htb]
    \centering
    \includegraphics[scale=1.0]{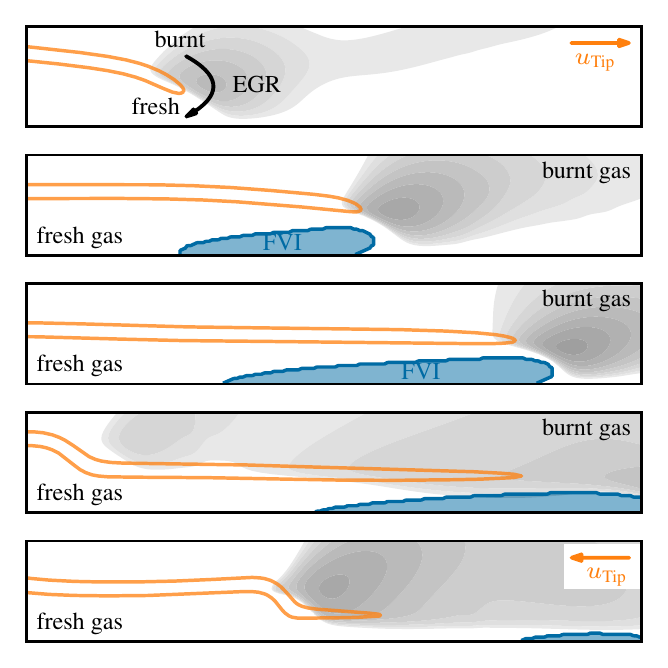}
    \caption{Annotated visualization of the FVI mechanism shown in Fig.~\ref{fig:time-series}. In orange, the flame front visualized by a contour of  $\dot{\omega}_\mathrm{CO_2}$ is shown, while the area of FVI is colored in blue. The vertical structures are shown in grey.}
    \label{fig:time-series-schematic}
\end{figure}

To assess this hypothesis, Fig.~\ref{fig:time-series} shows a time series for a slice of the flame in the lateral direction that displays a representative time series found in the turbulent SWQ flame. The area shown in the slice is marked by the blue box in Fig.~\ref{fig:DNS-setup}. On the left, the flame is visualized by the reaction rate of $\mathrm{CO_2}$, together with vortical structures indicated by the Q-criterion \cite{Hunt1988} (white lines). On the right, $Y_\mathrm{CO_2}$ is shown, as well as the reaction zone of the flame (black isoline of $\dot{\omega}_\mathrm{CO_2}=400~\text{1/s}$). Furthermore, the area where $Y_\mathrm{CO}$ falls below the conditional minimum of the laminar SWQ flame (Fig.~\ref{fig:thermochemical-state}(c)) is indicated by a white isoline. In the following, this area is referred to as the area of FVI. In Fig.~\ref{fig:time-series-schematic}, a schematic representation of the FVI mechanism is shown for the same area and time steps that are depicted in Fig.~\ref{fig:time-series}. 
At $t=0~\text{ms}$ the flame is in a SWQ-like state, and no areas of FVI are observed. A vortical structure is present downstream of the flame tip. The flame tip propagates into the vortical structure, pushing the vortex in the stream-wise direction. This forwards motion, together with the entrainment of hot exhaust gases at the wall, leads to the mixing of cold, burnt products and fresh gases below the reaction zone and thus between the flame branch and the cold wall (see $t=\left[0.55, 1.1\right]~\text{ms}$). The mixing changes the flame velocity at the flame tip, leading to decreased heat release at the flame tip ($t=1.65~\text{ms}$). At the same time, the vortex above the flame front spreads over a wide area and pushes the flame against the wall, resulting in a HOQ-like quenching region parallel to the wall ($t=2.2~\text{ms}$). After the flame has been pushed back, it spreads out again, and the FVI mechanism described is repeated. The observed flame behavior is in agreement with the experimental hypothesis by Zentgraf et al.~\cite{Zentgraf2021}
and is mainly driven by the interaction of the flame with the near-wall vortical structures. It is not restricted to the equivalence ratio or even fuel. A video of the temporal evolution of the flame can be found in the supplementary material, including two additional lateral positions.

\begin{figure}[!htb]
    \centering
    \includegraphics[scale=1.0]{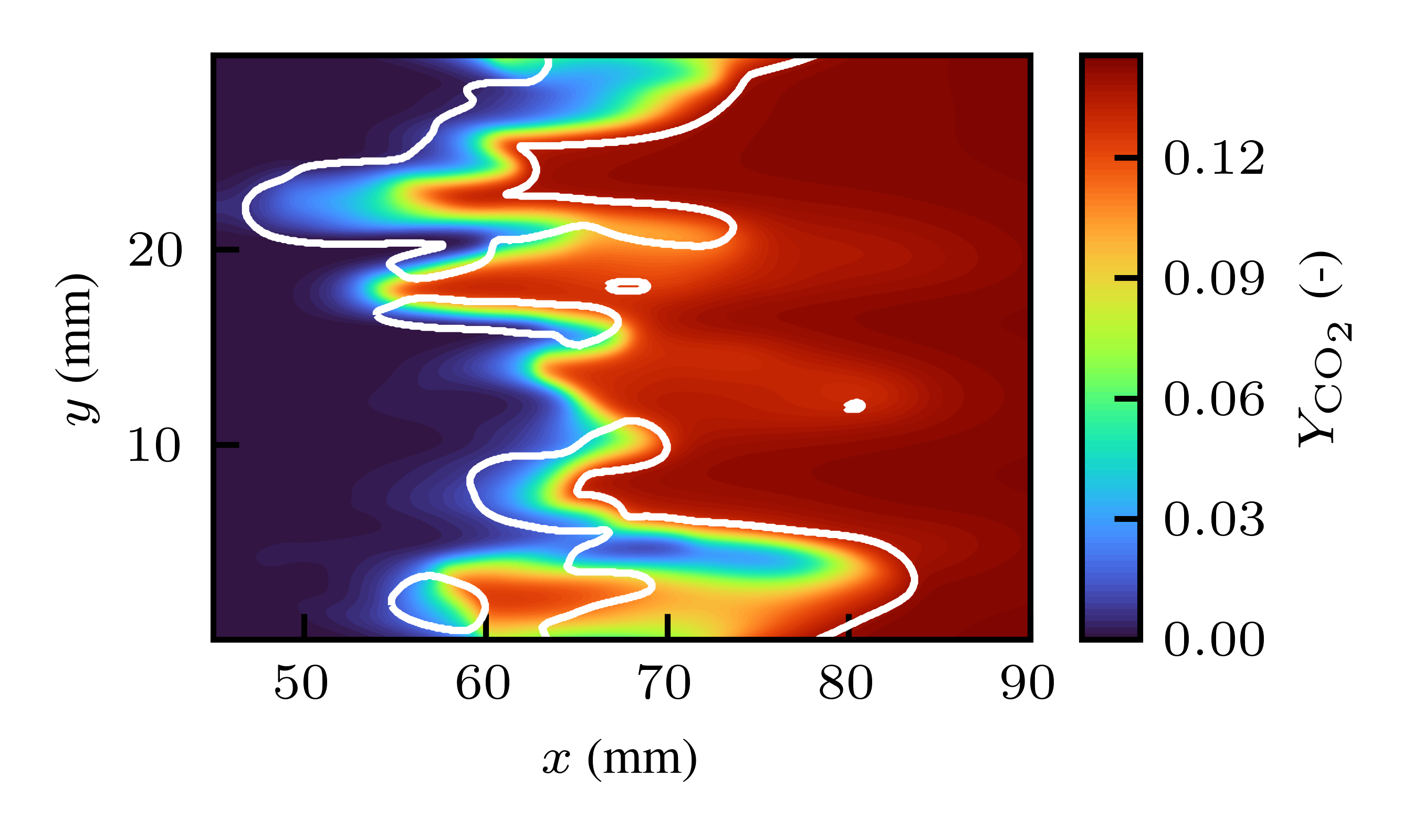}
    \caption{Wall-parallel cut through the simulation domain at $z/\delta_\text{L}=0.2$ and $t=1.1~\text{ms}$. The FVI area is indicated by the white isoline. The statistically independent lateral direction ($y$) can be understood as multiple realizations of the temporal flame evolution.}
    \label{fig:snapshot-FVI}
\end{figure}

To assess the importance of the FVI mechanism, Fig.~\ref{fig:snapshot-FVI} shows a wall-parallel cut through the simulation domain at $z/\delta_\text{L}=0.2$ and $t=1.1~\text{ms}$. The investigated flow configuration is statistically independent in the lateral direction. Therefore, the lateral direction can be understood as multiple realizations of the temporal flame evolution, and the wall-normal cut is a reasonable indication for the likelihood of the exhaust gas recirculation events caused by FVI at a given wall distance. In the figure, the area of FVI, indicated by the white isoline, is distributed over most of the reaction zone of the flame. A similar picture can also be observed at other time instances of the flame. With increasing distance from the wall, the area affected by FVI decreases, see also Fig.~\ref{fig:validate-lookup-statistic}. This shows the importance of FVI events in turbulent FWI, since in the near-wall region most of the reaction zone of the flame is prone to exhaust gas recirculation effects. Twenty simulation time steps throughout $5~\text{ms}$ have been analyzed concerning the probability of a FVI event to emphasize this aspect even further. Thereby, a FVI event was counted for every time step and lateral direction, if $\left( Y_\mathrm{CO} | Y_\mathrm{CO_2} \right)$ falls below the laminar SWQ simulation counterpart for more than one stream-wise location. In the analysis more than 80~\% of the lateral locations are prone to exhaust gas recirculation, confirming the observations made in the single time step shown in Fig.~\ref{fig:snapshot-FVI}. In the supplementary material the temporal evolution of three wall-normal slices at different wall distances are provided, additionally.

\section{Flamelet manifolds for turbulent SWQ} 
\begin{figure}[htb!]
\centering
\includegraphics[scale=1.0]{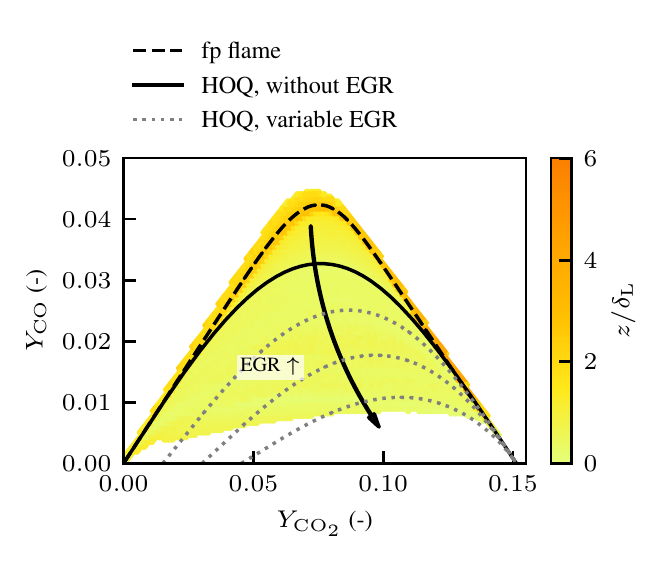}
\caption{Thermochemical states in the SWQ simulation colored by normalized wall distance $z/\delta_\text{L}$. Additionally, reference data from 1D Head-On Quenching (HOQ) simulations and a freely propagating (fp) flame is shown.}
\label{fig:thermochemical-state-EGR}
\end{figure}

\addvspace{10pt}
\label{sec:tabulation} 
Based on the above discussion, a novel manifold is proposed for turbulent SWQ as an extension of Quenching Flamelet-Generated Manifolds (QFMs)~\cite{Efimov2019b, Steinhausen2021}. In particular, an additional dimension is introduced to include the effects of exhaust gas recirculation (EGR) on the thermochemical state. The manifold consists of an ensemble of 1D HOQ flames. Starting from a freely propagating flame with fresh mixture on the unburnt side, the flame is gradually diluted with cold exhaust gases at inflow temperature. Each of the diluted flames is then used as the initial condition for a transient HOQ simulation with a wall temperature of 300~K. The numerical setup for the HOQ flames is described in detail in~\cite{Steinhausen2021}. In total, 63 HOQ flames are calculated with varying EGR levels from 0 to 0.3. Further, the upper boundary of the manifold is extended using preheated HOQ flames ranging from 300~K to 750~K. The ensemble of HOQ flames spans a 3D manifold in ($x$, $t$, $Y_\text{EGR}$) space, with $x$ being the spatial coordinate, $t$ the time, and $Y_\text{EGR}$ the amount of (cold) burnt gases mixed into the fresh gases. Figure~\ref{fig:thermochemical-state-EGR} shows the thermochemical state from the HOQ manifold with and without EGR compared against the turbulent SWQ. The diluted flamelets exhibit a lowered limit of $\left(Y_\mathrm{CO}|Y_\mathrm{CO_2}\right)$ and can cover the complete thermochemical state of the turbulent SWQ configuration. 

\newpage

The newly developed manifold is mapped in a three-step procedure onto a ($c$, $c_2$, $h_\text{norm}$) state with a resolution of ($150\times201\times101$). The variables $c$ and $c_2$ are normalized progress variables and $h_\text{norm}$ is the normalized enthalpy based on the total enthalpy $h$ as sum of sensible and enthalpy of formation
\begin{align}
\label{eq:normalized_PV1}
c&=\frac{Y_\text{c} - Y_\text{c,min}}{Y_\text{c,max}-Y_\text{c,min}} \ , \\
\label{eq:normalized_PV2}
c_2&=\frac{Y_\text{c,2} - Y_\text{c,2, min}\left( c \right)}{Y_\text{c,2,max}\left( c \right)-Y_\text{c,2,min}\left( c \right)} \ , \\
\label{eq:normalized_h}
h_\text{norm}&=\frac{h - h_\text{min} \left( c, c_2 \right)}{h_\text{max}\left( c, c_2 \right)-h_\text{min}\left( c, c_2 \right)} \ .
\end{align}

\noindent The mapped manifold can then be accessed in a three-step look-up with $c$ being the first, $c_2$ the second, and $h_\text{norm}$ the third look-up parameter. The first and second progress variable $Y_\text{c}$ and $Y_\text{c,2}$ are chosen to be the mass fraction of CO2 and CO, respectively. The final manifold is referred to as QFM-EGR (Quenching Flamelet-Generated Manifold with Exhaust Gas Recirculation).

\section{\textit{A-priori} validation of the extended manifold} 
\begin{figure*}[!htb]
    \centering
    \includegraphics[width=5.6in]{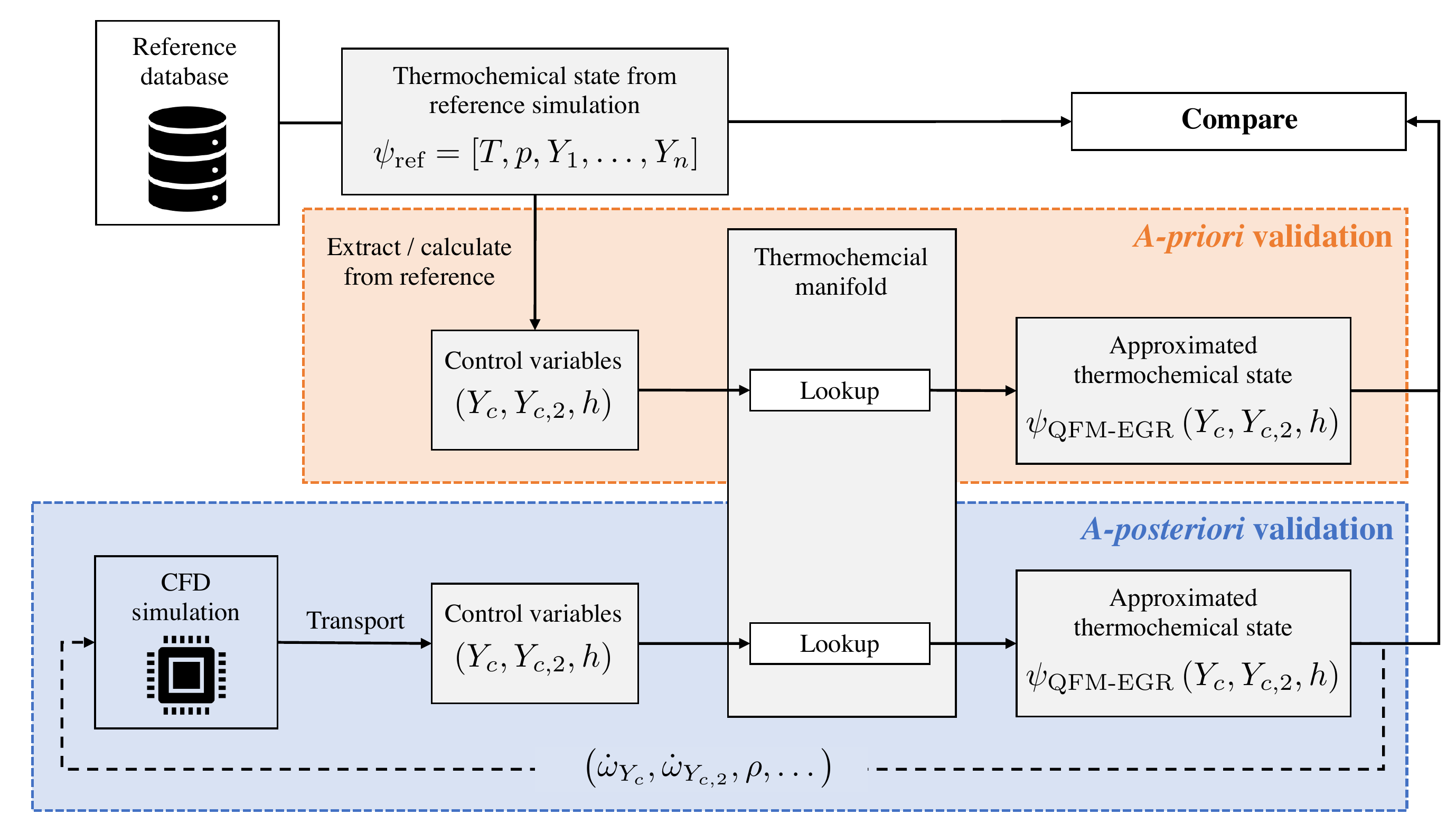}
    \vspace{10pt}
    \caption{Schematic representation of an \textit{a-priori} analysis in comparison to an \textit{a-posteriori} assessment in a fully coupled simulation. The approximated thermochemical states are compared to the reference data (here: turbulent SWQ simulation). The control variables shown in the figures correspond to the 3D QFM-EGR. For other manifolds the choice of control variables may differ. Note that in the CFD, the non-normalized progress variables are solved for. The normalized progress variables as shown in Eqs.~\eqref{eq:normalized_PV1}-\eqref{eq:normalized_h} are calculated during the lookup on the manifold.}
    \label{fig:a-priori}
\end{figure*}

\addvspace{10pt}
The extended 3D QFM-EGR is validated in an \textit{a-priori} analysis using the turbulent SWQ simulation data as a reference. Figure~\ref{fig:a-priori} shows the procedure of the \textit{a-priori} manifold assessment in comparison to a fully coupled (\textit{a-posteriori}) validation. While in the fully coupled simulation transport equations for the control variables of the manifold are solved, in the \textit{a-priori} validation, the control variables are taken directly from the reference data. This allows a detailed validation of the tabulated thermochemical state without the interference of errors caused by the inaccuracies in the solution of the transport equations in a coupled simulation.

\begin{figure}[!htb]
    \centering
    \includegraphics[width=192pt]{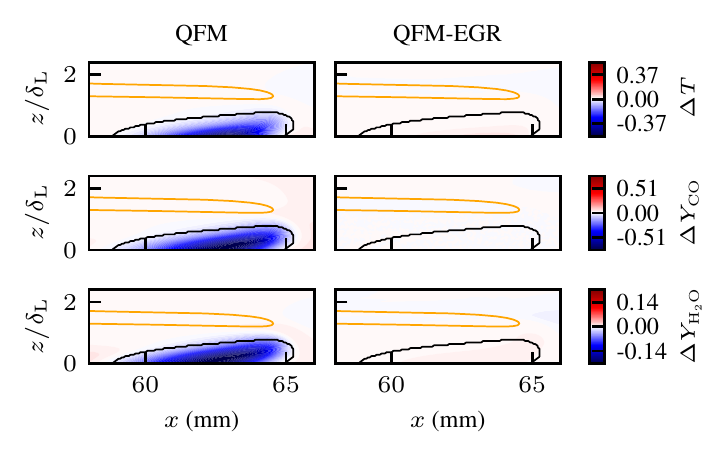}
    \caption{Normalized difference between \textit{a-priori} lookup quantities and the validation state at $t=1.10~\text{ms}$. The plot shows results from 2D QFM (left) and the novel QFM-EGR (right) derived in this work. Note that the figure shows a zoom of the relevant area depicted in Fig.~\ref{fig:time-series}.}
    \label{fig:validate-lookup}
\end{figure}

Figure~\ref{fig:validate-lookup} shows the results of the \textit{a-priori} validation for a QFM with two table dimensions and the extended three-dimensional QFM-EGR on the slice shown in Fig.~\ref{fig:time-series} at $t=1.10~\text{ms}$. The QFM consists of a single HOQ simulation of a freely propagating flame without EGR and, therefore, represents a subset of the QFM-EGR. The manifold generation of the QFM is described in more detail in~\cite{Steinhausen2021, Efimov2019b}. 
The color code shows the relative deviation of the approximated quantity $q$ from the reference data (turbulent SWQ simulation).

\begin{equation}
    \Delta q = \frac{q_\text{ref} - q_\text{Manifold}}{q_\text{ref}} \ .
\end{equation}

\noindent In accordance with Fig.~\ref{fig:time-series}, the flame front is visualized (orange contour), and the area of FVI is shown (black isoline). On the left, the lookup result for a QFM is shown, while on the right the QFM-EGR is depicted. Both manifolds show good agreement outside the area of FVI. Inside that area, however, the QFM shows discrepancies of up to 70~\% for all species studied, indicating that not only the mass fraction of CO is incorrectly predicted, but the mixture at the wall in the reference data is not consistent with the tabulated states in the QFM. This results in an incorrect prediction of the composition space as a whole. The new tabulation approach accounts for this shift in mixture caused by the EGR by introducing the additional table dimension $c_2$, leading to a very good agreement with the reference simulation. In addition to the data shown, further time series of the \textit{a-priori} validation are provided in the supplementary material for different lateral and wall-normal slices. The data also includes radicals and reaction rate predictions. The supplementary videos show the reference simulation and the prediction by the manifolds on the left, while the manifold deviations from the reference are depicted on the right.

\begin{figure}[!htb]
    \centering
    \includegraphics[width=192pt]{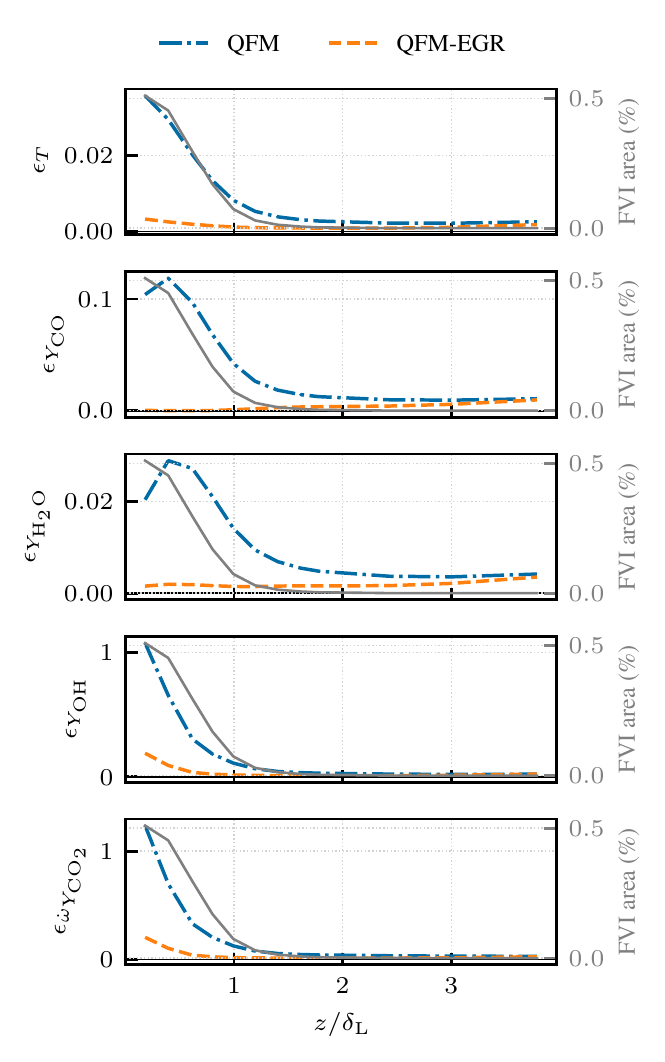}
    \caption{Mean relative error of the \textit{a-priori} lookup quantities and the reference simulation as a function of the normalized wall distance $z/\delta_\text{L}$.}
    \label{fig:validate-lookup-statistic}
\end{figure}

Finally, Fig.~\ref{fig:validate-lookup-statistic} shows the mean relative error for a quantity $q$
\begin{equation}
    \epsilon_q\left(z \right) = \frac{1}{N} \frac{ \sum_{\Omega\left(z\right)} \left| q_{i,\text{ref}} - q_{i,\text{Manifold}} \right| }{ \sum_{\Omega\left(z\right)} \left| q_{i,\text{ref}}\right| } \ ,
\end{equation}

\noindent with $N$ being the number of samples in $\Omega$ (reaction zone of the flame) and
\begin{equation}
    \Omega\left(z\right) = \lbrace(t,x,y,z) \vert c^*\in[0.3,0.7] \land z=z \rbrace \ ,
\end{equation}

\noindent with
\begin{equation}
    c^* = \frac{ Y_\mathrm{CO_2} - Y_\mathrm{CO_2,min}\left(h\right) }{ Y_\mathrm{CO_2,max}\left(h\right)-Y_\mathrm{CO_2,min}\left(h\right)}
\end{equation} 

\noindent being the normalized progress variable based on a given enthalpy. This definition was also used in~\cite{Steinhausen2022} to track the reaction zone of the flame. The time average is calculated using twenty simulation time steps throughout 5~ms. In addition to the quantities discussed above, the mass fraction of the OH-radical $Y_\text{OH}$ and the reaction rate of $\mathrm{CO_2}$ are depicted. Again, the QFM prediction capability worsens in the areas of FVI close to the wall. Outside the FVI area ($z/\delta_\mathrm{L}>2$), the prediction accuracy of the QFM is very good, only slight deviations are found for the CO and $\mathrm{H_2O}$ mass fraction. The QFM-EGR shows an excellent agreement with the reference solution. Note that the observed increase in prediction error of the radicals and reaction rates in the near-wall region is not due to an increase in the absolute prediction error, but due to the fact, that the value of both the OH-radical mass fraction and reaction rate approach zero due to the high enthalpy losses at the wall.

\addvspace{20pt}

\section{Conclusion} \addvspace{10pt}
In the present study, the thermochemical state of a turbulent methane-air flame interacting with a cold wall is investigated using a fully resolved simulation with detailed chemistry. The detailed insights from the simulation are used to confirm a recently proposed flame-vortex interaction mechanism, which describes the entrainment of burnt gases into the fresh gas mixture near the flame's quenching point.
Based on these findings, an extended flamelet manifold generated from an ensemble of 1D HOQ flames is presented, accounting for this particular mixing process. In the manifold, the effects of flame-vortex interaction on the thermochemical state are accounted for by introducing a new dimension to the manifold, which accounts for the shift in mixture caused by exhaust gas recirculation. The new manifold is validated in an \textit{a-priori} analysis. Without the additional dimension, large deviations between the previous manifold prediction (QFM) and the detailed reference simulation are observed in the flame-vortex interaction area near the cold wall. Accounting for exhaust gas recirculation in the manifold leads to significant improvements in the prediction of the thermochemical states. The improved manifold (QFM-EGR) constitutes a significant advance for the modeling of turbulent SWQ and will be further evaluated in coupled LES in future work.

\newpage

\section*{Acknowledgments} \addvspace{10pt}
This work has been funded by the Deutsche Forschungsgemeinschaft (DFG, German Research Foundation) -- Project Number 237267381 -- TRR 150 and the Center of Excellence in Combustion project, grant agreement No 952181 (F. Ferraro). The simulations were performed on the national supercomputer HAWK at the High Performance Computing Center Stuttgart (HLRS).

\bibliography{publication.bib}

\begin{thebibliography}{24}
\providecommand{\natexlab}[1]{#1}
\providecommand{\url}[1]{\texttt{#1}}
\expandafter\ifx\csname urlstyle\endcsname\relax
  \providecommand{\doi}[1]{doi: #1}\else
  \providecommand{\doi}{doi: \begingroup \urlstyle{rm}\Url}\fi

\bibitem[Poinsot and Veynante(2005)]{Poinsot2005a}
T.~Poinsot and D.~Veynante.
\newblock \emph{{Theoretical and Numerical Combustion}}.
\newblock R.T. Edwards, Inc., Philadelphia, USA, 2 edition, 2005.
\newblock ISBN 9781930217102.

\bibitem[Fritz et~al.(2004)Fritz, Kr{\"{o}}ner, and Sattelmayer]{Fritz2004}
J.~Fritz, M.~Kr{\"{o}}ner, and T.~Sattelmayer.
\newblock {Flashback in a Swirl Burner With Cylindrical Premixing Zone}.
\newblock \emph{J. Eng. Gas Turbines Power}, 126\penalty0 (2):\penalty0
  276--283, apr 2004.
\newblock \doi{10.1115/1.1473155}.

\bibitem[Gruber et~al.(2010)Gruber, Sankaran, Hawkes, and Chen]{Gruber2010}
A.~Gruber, R.~Sankaran, E.~R. Hawkes, and J.~H. Chen.
\newblock {Turbulent flame-wall interaction: A direct numerical simulation
  study}.
\newblock \emph{J. Fluid Mech.}, 658:\penalty0 5--32, 2010.
\newblock \doi{10.1017/S0022112010001278}.

\bibitem[Ahmed et~al.(2021)Ahmed, Chakraborty, and Klein]{Ahmed2021}
U.~Ahmed, N.~Chakraborty, and M.~Klein.
\newblock {Scalar Gradient and Strain Rate Statistics in Oblique Premixed
  Flame-Wall Interaction Within Turbulent Channel Flows}.
\newblock \emph{Flow, Turbul. Combust.}, 106\penalty0 (2):\penalty0 701--732,
  feb 2021.
\newblock \doi{10.1007/s10494-020-00169-3}.

\bibitem[Steinhausen et~al.(2022)Steinhausen, Zirwes, Ferraro, Popp, Zhang,
  Bockhorn, and Hasse]{Steinhausen2022}
M.~Steinhausen, T.~Zirwes, F.~Ferraro, S.~Popp, F.~Zhang, H.~Bockhorn, and
  C.~Hasse.
\newblock {Turbulent flame-wall interaction of premixed flames using
  Quadrature-based Moment Methods (QbMM) and tabulated chemistry: An a priori
  analysis}.
\newblock \emph{Int. J. Heat Fluid Flow}, 93\penalty0 (2):\penalty0 108913,
  2022.
\newblock \doi{10.1016/j.ijheatfluidflow.2021.108913}.

\bibitem[Jiang et~al.(2021)Jiang, Brouzet, Talei, Gordon, Cazeres, and
  Cuenot]{Jiang2021a}
B.~Jiang, D.~Brouzet, M.~Talei, R.~L. Gordon, Q.~Cazeres, and B.~Cuenot.
\newblock {Turbulent flame-wall interactions for flames diluted by hot
  combustion products}.
\newblock \emph{Combust. Flame}, 230\penalty0 (8):\penalty0 111432, aug 2021.
\newblock \doi{10.1016/j.combustflame.2021.111432}.

\bibitem[van Oijen and de~Goey(2000)]{VanOijen2000}
J.~A. van Oijen and L.~P.~H. de~Goey.
\newblock {Modelling of Premixed Laminar Flames using Flamelet-Generated
  Manifolds}.
\newblock \emph{Combust. Sci. Technol.}, 161\penalty0 (1):\penalty0 113--137,
  2000.
\newblock \doi{10.1080/00102200008935814}.

\bibitem[Gicquel et~al.(2000)Gicquel, Darabiha, and
  Th{\'{e}}venin]{Gicquel2000}
O.~Gicquel, N.~Darabiha, and D.~Th{\'{e}}venin.
\newblock {Laminar premixed hydrogen/air counterflow flame simulations using
  flame prolongation of ILDM with differential diffusion}.
\newblock \emph{Proc. Combust. Inst.}, 28\penalty0 (2):\penalty0 1901--1908,
  2000.
\newblock \doi{10.1016/s0082-0784(00)80594-9}.

\bibitem[Maas and Pope(1992)]{Maas1992}
U.~Maas and S.~B. Pope.
\newblock {Simplifying chemical kinetics: Intrinsic low-dimensional manifolds
  in composition space}.
\newblock \emph{Combust. Flame}, 88:\penalty0 239--264, 1992.

\bibitem[Bykov and Maas(2007)]{Bykov2007}
V.~Bykov and U.~Maas.
\newblock {The extension of the ILDM concept to reaction-diffusion manifolds}.
\newblock \emph{Combust. Theory Model.}, 11\penalty0 (6):\penalty0 839--862,
  2007.
\newblock \doi{10.1080/13647830701242531}.

\bibitem[Ganter et~al.(2018)Ganter, Stra{\ss}acker, Kuenne, Meier, Heinrich,
  Maas, and Janicka]{Ganter2018}
S.~Ganter, C.~Stra{\ss}acker, G.~Kuenne, T.~Meier, A.~Heinrich, U.~Maas, and
  J.~Janicka.
\newblock {Laminar near-wall combustion: Analysis of tabulated chemistry
  simulations by means of detailed kinetics}.
\newblock \emph{Int. J. Heat Fluid Flow}, 70:\penalty0 259--270, 2018.
\newblock \doi{10.1016/j.ijheatfluidflow.2018.02.015}.

\bibitem[Strassacker et~al.(2021)Strassacker, Bykov, and Maas]{Strassacker2021}
C.~Strassacker, V.~Bykov, and U.~Maas.
\newblock {Comparative analysis of Reaction-Diffusion Manifold based reduced
  models for Head-On- and Side-Wall-Quenching flames}.
\newblock \emph{Proc. Combust. Inst.}, 38\penalty0 (1):\penalty0 1025--1032,
  jan 2021.
\newblock \doi{10.1016/J.PROCI.2020.06.130}.

\bibitem[Steinhausen et~al.(2021)Steinhausen, Luo, Popp, Strassacker, Zirwes,
  Kosaka, Zentgraf, Maas, Sadiki, Dreizler, and Hasse]{Steinhausen2021}
M.~Steinhausen, Y.~Luo, S.~Popp, C.~Strassacker, T.~Zirwes, H.~Kosaka,
  F.~Zentgraf, U.~Maas, A.~Sadiki, A.~Dreizler, and C.~Hasse.
\newblock {Numerical Investigation of Local Heat-Release Rates and
  Thermo-Chemical States in Side-Wall Quenching of Laminar Methane and Dimethyl
  Ether Flames}.
\newblock \emph{Flow, Turbul. Combust.}, 106\penalty0 (2):\penalty0 681--700,
  feb 2021.
\newblock \doi{10.1007/s10494-020-00146-w}.

\bibitem[Efimov et~al.(2020)Efimov, de~Goey, and van Oijen]{Efimov2019b}
D.~V. Efimov, P.~de~Goey, and J.~A. van Oijen.
\newblock {QFM: quenching flamelet-generated manifold for modelling of
  flame–wall interactions}.
\newblock \emph{Combust. Theory Model.}, 24\penalty0 (1):\penalty0 72--104, aug
  2020.
\newblock \doi{10.1080/13647830.2019.1658901}.

\bibitem[Kosaka et~al.(2018)Kosaka, Zentgraf, Scholtissek, Bischoff,
  H{\"{a}}ber, Suntz, Albert, Hasse, and Dreizler]{Kosaka2018}
H.~Kosaka, F.~Zentgraf, A.~Scholtissek, L.~Bischoff, T.~H{\"{a}}ber, R.~Suntz,
  B.~Albert, C.~Hasse, and A.~Dreizler.
\newblock {Wall heat fluxes and CO formation/oxidation during laminar and
  turbulent side-wall quenching of methane and DME flames}.
\newblock \emph{Int. J. Heat Fluid Flow}, 70\penalty0 (1):\penalty0 181--192,
  2018.
\newblock \doi{10.1016/j.ijheatfluidflow.2018.01.009}.

\bibitem[Zentgraf et~al.(2022)Zentgraf, Johe, Steinhausen, Hasse, Greifenstein,
  Cutler, Barlow, and Dreizler]{Zentgraf2021a}
F.~Zentgraf, P.~Johe, M.~Steinhausen, C.~Hasse, M.~Greifenstein, A.~D. Cutler,
  R.~S. Barlow, and A.~Dreizler.
\newblock {Detailed assessment of the thermochemistry in a side-wall quenching
  burner by simultaneous quantitative measurement of CO2, CO and temperature
  using laser diagnostics}.
\newblock \emph{Combust. Flame}, 235\penalty0 (1):\penalty0 111707, 2022.
\newblock \doi{10.1016/j.combustflame.2021.111707}.

\bibitem[Palulli et~al.(2019)Palulli, Talei, and Gordon]{Palulli2019}
R.~Palulli, M.~Talei, and R.~L. Gordon.
\newblock {Unsteady flame–wall interaction: Impact on CO emission and wall
  heat flux}.
\newblock \emph{Combust. Flame}, 207:\penalty0 406--416, sep 2019.
\newblock \doi{10.1016/J.COMBUSTFLAME.2019.06.012}.

\bibitem[Zentgraf et~al.(2021)Zentgraf, Johe, Cutler, Barlow, B{\"{o}}hm, and
  Dreizler]{Zentgraf2021}
F.~Zentgraf, P.~Johe, A.~D. Cutler, R.~S. Barlow, B.~B{\"{o}}hm, and
  A.~Dreizler.
\newblock {Classification of flame prehistory and quenching topology in a
  side-wall quenching burner at low-intensity turbulence by correlating
  transport effects with CO2, CO and temperature}.
\newblock \emph{Combust. Flame}, page 111681, 2021.
\newblock \doi{10.1016/j.combustflame.2021.111681}.

\bibitem[Jainski et~al.(2017)Jainski, Ri{\ss}mann, B{\"{o}}hm, and
  Dreizler]{Jainski2017a}
C.~Jainski, M.~Ri{\ss}mann, B.~B{\"{o}}hm, and A.~Dreizler.
\newblock {Experimental investigation of flame surface density and mean
  reaction rate during flame-wall interaction}.
\newblock \emph{Proc. Combust. Inst.}, 36\penalty0 (2):\penalty0 1827--1834,
  2017.
\newblock \doi{10.1016/j.proci.2016.07.113}.

\bibitem[Zirwes et~al.(2021)Zirwes, H{\"{a}}ber, Zhang, Kosaka, Dreizler,
  Steinhausen, Hasse, Stagni, Trimis, Suntz, and Bockhorn]{Zirwes2021}
T.~Zirwes, T.~H{\"{a}}ber, F.~Zhang, H.~Kosaka, A.~Dreizler, M.~Steinhausen,
  C.~Hasse, A.~Stagni, D.~Trimis, R.~Suntz, and H.~Bockhorn.
\newblock {Numerical Study of Quenching Distances for Side-Wall Quenching Using
  Detailed Diffusion and Chemistry}.
\newblock \emph{Flow, Turbul. Combust.}, 106\penalty0 (2):\penalty0 649--679,
  feb 2021.
\newblock \doi{10.1007/s10494-020-00215-0}.

\bibitem[Ranzi et~al.(2012)Ranzi, Frassoldati, Grana, Cuoci, Faravelli, Kelley,
  and Law]{Ranzi2012}
E.~Ranzi, A.~Frassoldati, R.~Grana, A.~Cuoci, T.~Faravelli, A.~P. Kelley, and
  C.~K. Law.
\newblock {Hierarchical and comparative kinetic modeling of laminar flame
  speeds of hydrocarbon and oxygenated fuels}, aug 2012.

\bibitem[Zirwes et~al.(2018{\natexlab{a}})Zirwes, Zhang, Denev, Habisreuther,
  and Bockhorn]{Zirwes2018}
T.~Zirwes, F.~Zhang, J.~A. Denev, P.~Habisreuther, and H.~Bockhorn.
\newblock {Automated code generation for maximizing performance of detailed
  chemistry calculations in OpenFOAM}.
\newblock In \emph{High Perform. Comput. Sci. Eng. 17 Trans. High Perform.
  Comput. Center, Stuttgart 2017}, pages 189--204. Springer International
  Publishing, jan 2018{\natexlab{a}}.
\newblock ISBN 9783319683942.
\newblock \doi{10.1007/978-3-319-68394-2_11}.

\bibitem[Zirwes et~al.(2018{\natexlab{b}})Zirwes, Zhang, Denev, Habisreuther,
  Bockhorn, and Trimis]{Zirwes2018improved}
T.~Zirwes, F.~Zhang, J.~A. Denev, P.~Habisreuther, H.~Bockhorn, and D.~Trimis.
\newblock {Improved Vectorization for Efficient Chemistry Computations in
  OpenFOAM for Large Scale Combustion Simulations}.
\newblock In W.~E. Nagel, D.~H. Kr{\"{o}}ner, and M.~M. Resch, editors,
  \emph{High Perform. Comput. Sci. Eng. '18}, pages 209--224. Springer,
  2018{\natexlab{b}}.
\newblock \doi{10.1007/978-3-030-13325-2_13}.

\bibitem[Hunt et~al.(1988)Hunt, Wray, and Moin]{Hunt1988}
J.~Hunt, A.~Wray, and P.~Moin.
\newblock {Eddies, Steams, and Convergence Zones in Turbulent Flows}.
\newblock In \emph{Cent. Turbul. Res. Proc. Summer Progr.}, pages 193--208,
  1988.

\end{thebibliography}
\bibliographystyle{unsrtnat_mod}

\end{document}